\documentclass[9pt,twocolumn,twoside]{osajnl}
\journal{ol} % Choose journal (ao,jocn,josaa,josab,ol,optica,pr)

\setboolean{shortarticle}{true}
\usepackage{lineno}
\title{On-axis polarization of beams radiated by electromagnetic circularly coherent sources}

\author[1]{J. C. G. De Sande}
\author[2,*]{O. Korotkova}
\author[3]{M. Santarsiero}
\author[4]{R. Mart\'{i}nez-Herrero}
\author[4]{G. Piquero}
\author[3]{F. Gori}

\affil[1]{ETSIS de Telecomunicación, Universidad Polit\'ecnica de Madrid, Campus Sur 28031 Madrid, Spain}
\affil[2]{Department of Physics, University of Miami, 1320 Campo Sano Drive, Coral Gables, FL, 33146}
\affil[3]{Dipartimento di Ingegneria Industriale, Elettronica e Meccanica, Universit\'a Roma Tre, Via V. Volterra 62, 00146 Rome, Italy}
\affil[4]{Departamento de \'Optica, Universidad Complutense de Madrid, Ciudad Universitaria, 28040 Madrid, Spain}

\affil[*]{Corresponding author: korotkova@physics.miami.edu
\newline
\newline
\bf © 2022 Optica Publishing Group. One print or electronic copy may be made for personal use only. Systematic reproduction and distribution, duplication of any material in this paper for a fee or for commercial purposes, or modifications of the content of this paper are prohibited.
\newline
\url{https://doi.org/10.1364/OL.465816}
}

\begin{abstract}
On-axis spectral density and degree of polarization of beams radiated by electromagnetic (EM) sources with circular correlations are shown to be finely controlled by changing the source parameters. We reveal, in particular, that in this beam class, unlike for all  previously known stationary beams, it is possible to control independently the dynamics of the on-axis spectral density and the degree of polarization. This was enabled by the obtained analytical expression for the on-axis polarization matrix, derived for general EM sources with circular coherence and Gaussian spectral density across the source plane. A simple experimental scheme for generating a broad class of  EM circularly coherent sources is devised involving only a line source, a lens and a  transparency, possibly anisotropic.  
\end{abstract}

\setboolean{displaycopyright}{true}

\begin{document}

\maketitle

The ability of stationary light sources to radiate beam-like fields with the degree and the state of polarization changing on propagation in vacuum has been illustrated a long time ago \cite{James:94, Agrawal:JOSAA00, Gori_2000, Piquero:01, KOROTKOVA2005, Salem:OL08, Vidal:PRA11, Santarsiero:JOPT13}. These early studies were based on Schell-model (uniformly correlated) sources and were shown to exhibit drastic changes in the on-axis polarization properties. Later, the electromagnetic (EM) non-uniformly correlated sources and those with the phase twist illustrated even more interesting dynamics in the on-axis polarization \cite{Tong:12, Mei:18, Zhao:OL18, Zhao:OL19, Joshi:OL20, Hyde:Opt20}.      
The recent findings by the authors relating to beams radiated by sources with circular coherence \cite{Santarsiero:OL17,Santarsiero:OL17b} confirmed that the fine spectral density control is readily achievable along the optical axis \cite{zcoh:OL2021,OK:22}. The aim of this Letter is to generalize the main analytical relations obtained in \cite{OK:22} from scalar to EM domain and illustrate the possibility of fine control in the on-axis degree of polarization's dynamics, whether or not synchronized with that in the spectral density.  

Scalar sources with circular coherence are characterized by a Cross-Spectral Density (CSD) of the form~\cite{OK:22}
\begin{equation}
W_0(\pmb{\rho}_1,\pmb{\rho}_2)
=
\tau(\pmb{\rho}_1) \tau^*(\pmb{\rho}_2) 
\; g(\rho^2_2-\rho^2_1)
\; ,
\label{CCscalar}
\end{equation}
where the $\ast$ denotes complex conjugate, $\tau$ is an arbitrary function of the coordinate across the source plane ($\pmb{\rho}$) and is related to the intensity profile of the latter, while $g$ is a function that determines the coherence properties of the source. Such sources exhibit perfect coherence along any annulus that is concentric to the source center, while coherence can be partial or even vanishing between two points at different distances from the center.

For the case of a vector source, the self and the joint second-order spatial correlations of the two mutually orthogonal components of the electric field $\overrightarrow{E}$ at two points of the source plane ($z=0$) are characterized by the $2\times 2$ cross-spectral density matrix (CSDM)~\cite{Wolf07}, i.e., 
\begin{equation}
\overleftrightarrow{W}^{(0)}(\pmb{\rho}_1,\pmb{\rho}_{2})
=
\langle \overrightarrow{E}(\pmb{\rho}_1)\overrightarrow{E}^{\dagger}(\pmb{\rho}_2)\rangle,    
\label{CSDM}
\end{equation}
where angular brackets stand for ensemble average and dagger denotes Hermitian adjoint. The natural extension of the CSD of Eq.~(\ref{CCscalar}) to the vector case is a CSDM of the form
%We will assume that the CSDM has the Gaussian spectral density,
\begin{equation}
\label{CSDMcirc}
\overleftrightarrow{W}^{(0)}(\pmb{\rho}_1,\pmb{\rho}_{2})
=
%\exp\left[-\frac{\pmb{\rho}_1^2+\pmb{\rho}_2^2}{w_0^2}\right]
\overleftrightarrow{T}(\pmb{\rho}_1)
\overleftrightarrow{G}^{(0)}\left(\rho_2^2-\rho_1^2 \right)
\overleftrightarrow{T}^{\dagger}(\pmb{\rho}_2),
\end{equation}
where $\overleftrightarrow{T}(\pmb{\rho})$ is a general $2 \times 2$ matrix, and $\overleftrightarrow{G}^{(0)}$ a $2\times 2$ matrix giving account of the correlations among all the field components. We shall loosely refer to $\overleftrightarrow{G}^{(0)}$ as the degree of coherence matrix across the source plane. The local polarization features of the source are accounted for by the polarization matrix
\begin{equation}
\label{Polarcirc}
\overleftrightarrow{S}^{(0)}(\pmb{\rho})
=
\overleftrightarrow{W}^{(0)}(\pmb{\rho},\pmb{\rho})
=
\overleftrightarrow{T}(\pmb{\rho})
\overleftrightarrow{G}^{(0)}(0)
\overleftrightarrow{T}^{\dagger}(\pmb{\rho})
\; .
\end{equation}

The realizability conditions on the matrices appearing in Eq.~(\ref{CSDMcirc}) can be found from the vectorial form of the Parzen integral representation of the CSDM \cite{Gori:2009, MH:2009} according to which a CSDM is well defined if it can expressed as
\begin{equation}
\label{Bochner}
\overleftrightarrow{W}^{(0)}(\pmb{\rho}_1,\pmb{\rho}_{2})
= 
\int \limits_{-\infty}^{\infty} \overleftrightarrow{H}(\pmb{\rho}_1,v) \overleftrightarrow{p}(v) \overleftrightarrow{H}^{\dagger}(\pmb{\rho}_2, v) \; {\rm d}v,
\end{equation}
where $v$ is a scalar parameter, $\overleftrightarrow{p}$ is a $2\times 2$ Hermitian, non-negative definite  matrix for any value of $v$, i.e.,
\begin{equation}\label{conditions}
\begin{array}{c}
p_{xy}(v)=p_{yx}^{\ast}(v),   \\
p_{xx}(v) \geq 0, \quad p_{yy}(v) \geq 0, \quad
\text{det}\overleftrightarrow{p}(v) \geq 0,
\end{array}
\end{equation}
det standing for matrix determinant, and $\overleftrightarrow{H}$ is $2\times 2$ matrix. The CSDM in \eqref{CSDMcirc} can be obtained for $\overleftrightarrow{H}$ selected as
\begin{equation}
\label{H}
    \overleftrightarrow{H}(\pmb{\rho},v)
    =
    \overleftrightarrow{T}(\pmb{\rho})
    \;
    e^{{\rm i} \,  v \rho^2}
%    \exp\left[-{\rm i} v \rho^2\right]
    \; ,
\end{equation}
which, if used in Eq.~(\ref{Bochner}), gives the same form as in Eq.~(\ref{CSDMcirc}) with 
\begin{equation}
\label{mucirc}
\overleftrightarrow{G}^{(0)} \left(\rho_1^2-\rho_2^2 \right)
=
\int \limits_{-\infty}^{\infty} \overleftrightarrow{p}(v) \exp\left[ - {\rm i} \,  \left(\rho_2^2-\rho_1^2 \right) v \right]  {\rm d}v
\; .
\end{equation}

The latter expression is the Fourier transform of the matrix $ \overleftrightarrow{p}$ evaluated at $(\rho_2^2-\rho_1^2)$.  Therefore, the one in Eq.~(\ref{CSDMcirc}) is a \emph{bona-fide} CSDM for any choice of $\overleftrightarrow{T}(\pmb{\rho})$, provided that the Fourier transform of  $\overleftrightarrow{G}^{(0)}$ is a Hermitian, non-negative definite matrix.

Among the possible choices of the matrices $\overleftrightarrow{T}(\pmb{\rho})$ and $\overleftrightarrow{G}^{(0)}$ giving rise to the CSDM of a well defined vector source with circular coherence we take 
\begin{equation}
\label{Tmatrix}
\overleftrightarrow{T}(\pmb{\rho})
=
\begin{bmatrix}
 A_x & 0 \\ 0 & A_y 
 \end{bmatrix} 
 \exp\left[-\frac{\rho^2}{w_0^2}\right] 
    \; ,
\end{equation}
where $w_0$, $A_x$, and $A_y$ are real quantities. Furthermore, we consider correlations among the field components that have all the same functional form (say, $\mu$) but might differ in characteristic widths. It will then be convenient to express the scalar correlations in a scaled form and set, for the elements of $\overleftrightarrow{G}^{(0)}$,
\begin{equation}
G_{ij}^{(0)}\left(\rho_2^2-\rho_1^2\right)
=
\beta_{ij} \; \mu\left(\frac{\rho_2^2-\rho_1^2}{\delta_{ij}^2}\right)
\;  ,
\end{equation}
where $(i,j)=(x,y)$, $\delta_{ij}$ are the typical correlation widths, and $\beta_{ij}$ are single-point generally complex-valued correlations. Thus, the elements of the resulting CSDM acquire the form
\begin{equation} \label{Wijcirc}
W^{(0)}_{ij}(\pmb{\rho}_1,\pmb{\rho}_{2})=A_iA_j\beta_{ij} \exp\left[-\frac{\rho_1^2+\rho_2^2}{w_0^2}\right] \mu\left(\frac{\rho_2^2-\rho_1^2}{\delta_{ij}^2}\right).
\end{equation}
From the condition $W^{(0)}_{xy}(\pmb{\rho}_1,\pmb{\rho}_{2})=W^{(0)\ast}_{yx}(\pmb{\rho}_2,\pmb{\rho}_{1})$,
it follows that $\delta_{ij}=\delta_{ji}$ and $\beta_{ij}=\beta^*_{ji}$.
For simplicity, we choose $\beta_{xx}=\beta_{yy}=1$ so that, from non-negativeness of $\overleftrightarrow{p}$ matrix, 
$|\beta_{xy}|\leq 1$. %,  $|\beta_{yx}|\leq 1$.

Then, on introducing the variables
\begin{equation}
q_{ij}=w_0^2/\delta_{ij}^2, \quad \zeta=z/z_R, \quad z_R=\pi w_0^2/\lambda,
\end{equation}
we find, on {straightforward generalization of} Eq. (11) in \cite{OK:22} {from a scalar on-axis spectral density $S(\zeta)$ to the four components of the on-axis CSDM $S_{ij}(\zeta)$,} that 
%the on-axis single-point components of the CSDM, say $\overleftrightarrow{S}(\zeta)=\overleftrightarrow{W}(\zeta,\zeta)$, become
\begin{equation}\label{Sz}
S_{ij}(\zeta)=\frac{A_iA_j\beta_{ij}}{q_{ij}\zeta^2} \text{Re}\left\{ \widehat {\mu}\left( \frac{\zeta-i}{q_{ij} \zeta}  \right)  \right\},
\end{equation}
where $\widehat{\mu}$ denotes the Laplace transform of $\mu$~\cite{Abramowitz72}.

The on-axis spectral density and the degree of polarization can be found analytically as \cite{Korotkova2021a}
\begin{equation}\label{SP}
S(\zeta)= \text{tr} \overleftrightarrow{S}(\zeta), \quad   
P(\zeta)=\sqrt{1-\frac{4\text{det} \overleftrightarrow{S}(\zeta)}{\text{tr}^2 \overleftrightarrow{S}(\zeta)}},    
\end{equation}
tr denoting trace  of a matrix. Substitution from \eqref{Sz} into \eqref{SP} enables the analysis of the on-axis dynamics of the spectral density and the degree of polarization for circularly coherent sources. 

Consider now an example involving the Gaussian pair:
\begin{equation}
\label{G}
\begin{split}
&\mu\left(\frac{\rho_2^2-\rho_1^2}{\delta_{ij}^2} \right) = \exp \left[-\frac{(\rho_2^2-\rho_1^2)^2}{\delta_{ij}^4}\right], \\&
p_{ij}(v)= \frac{\beta_{ij}\delta_{ij}^2}{2\sqrt{\pi}} \exp\left(-\delta_{ij}^4v^2/4\right).
\end{split}
\end{equation}

It immediately follows from inequalities in  \eqref{conditions} that
\begin{equation}\label{rc_delta}
\frac{1}{2}(\delta_{xx}^4+\delta_{yy}^4)
\leq \delta_{xy}^4 \leq \frac{\delta_{xx}^2\delta_{yy^2}}{|\beta_{xy}|^2}.
\end{equation}
To derive these inequalities it sufficed use the monotonic behavior of Gaussian functions, setting $v=0$ and $v \rightarrow \infty$.

Note that, on using relations between $q_{ij}$ and $\delta_{ij}$, $(i,j)=(x,y)$, we also find that  %$\delta_{xx}\geq 0$, $\delta_{yy} \geq 0$, and 
\begin{equation}\label{rc_q}
%q_{xx}\geq 0, \quad q_{yy} \geq 0, \quad 
q_{xx}q_{yy}|\beta_{xy}|^2
\leq q_{xy}^2 \leq \frac{2q_{xx}^2q_{yy}^2}{{q_{xx}^2+q_{yy}^2}}.
\end{equation}

If the source degree of coherence has a Gaussian circular profile, as in \eqref{G}, the spectral density components in \eqref{Sz} take on the following simple analytic formula \cite{OK:22}:
\begin{equation}
\label{Sij}
 S_{ij}(\zeta)=\frac{\sqrt{\pi}A_iA_j\beta_{ij}}{2q_{ij}\zeta^2} \text{Re} \left\{\exp\left[  \frac{(\zeta-i)^2}{4q_{ij}^2\zeta^2}   \right]\text{ercf}\left(\frac{\zeta-i}{2q_{ij}\zeta} \right)\right\},     
\end{equation}
where \text{erfc} stands for the complimentary error function~\cite{Abramowitz72}. Note that since $\beta_{ij}$ are complex-valued for $i \neq j$, the same holds for $S_{ij}$.
Using such expressions in the definitions in \eqref{SP}, the on-axis properties of the radiated  EM beam can be studied. 

\begin{figure}[t]
\centering
\includegraphics[width=0.49\linewidth]{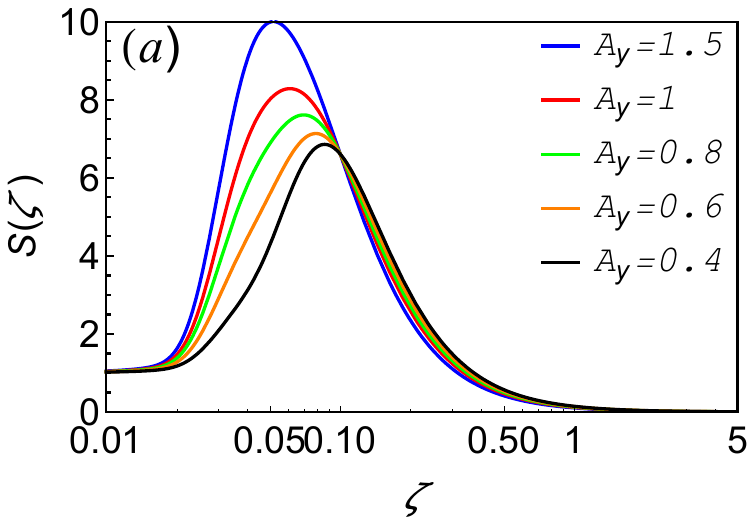}
\includegraphics[width=0.49\linewidth]{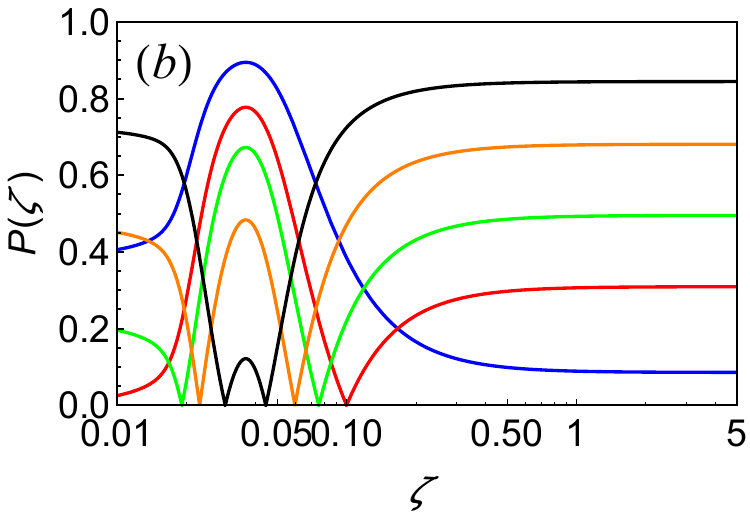}
\caption{(a) $S(\zeta)$; (b) $P(\zeta)$ of a beam radiated from a source given by \eqref{Wijcirc} with  $q_{xx}$ = 5, $q_{yy}$ = 10; $\beta_{xy}=0$, $A_{x}=1$; and several values of $A_{y}$.}
\label{fig1}
\end{figure}

Figure \ref{fig1} shows the behavior vs $\zeta$ of the on-axis spectral density and degree of polarization of the beam radiated from a circularly coherent source with uncorrelated $x$ and $y$ components ($\beta_{xy}=0$). In particular, it can be noted that $P$ generally presents a relative maximum and, depending on the cases, none, one or two zeros along the axis. Several maxima and zeros of the degree of polarization along the propagation axis have been reported for periodic structures~\cite{Piquero:01, Santarsiero:JOPT13}.

\begin{figure}[t]
\centering
\includegraphics[width=1\linewidth]{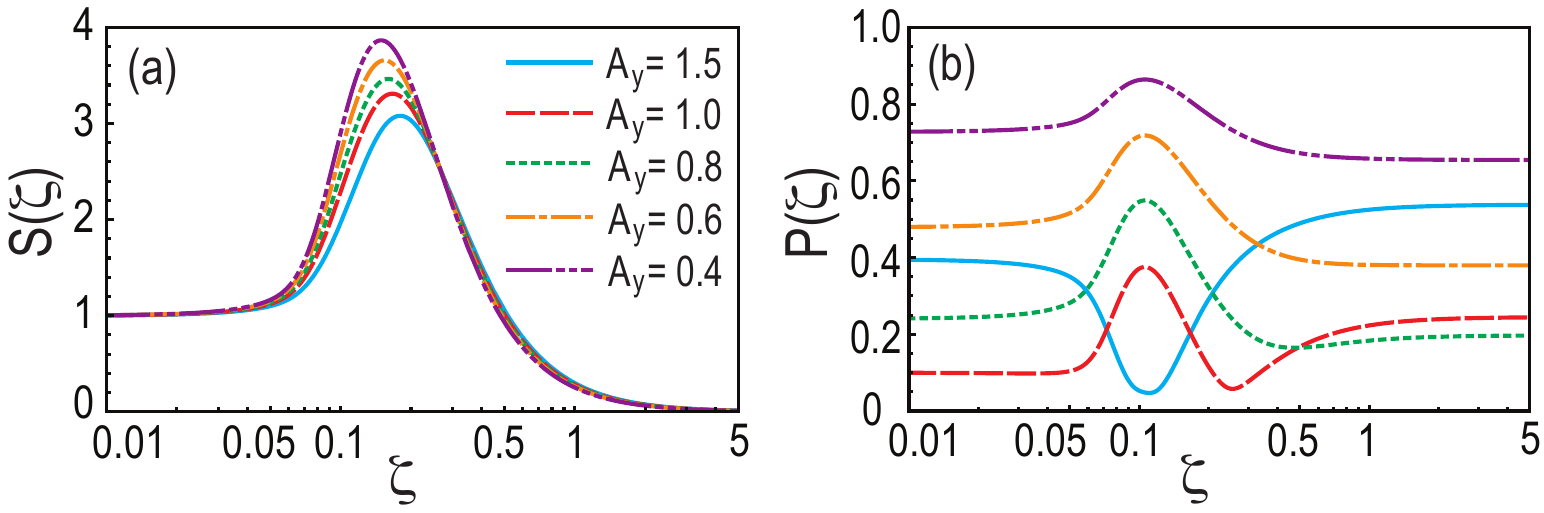}
\caption{(a) $S(\zeta)$; (b) $P(\zeta)$ of a beam radiated from a source given by \eqref{Wijcirc} with  $q_{xx} = 3$, $q_{yy} = 2 $; $q_{xy}=1$ $\beta_{xy}=0.1$, $A_{x}=1$; and several values of $A_{y}$.}
\label{fig2}
\end{figure}

Figure \ref{fig2} has the same structure as Fig. \ref{fig1} but describes a beam radiated by a source with $\beta_{xy}=0.1$. In this case $P$ never vanishes and presents a relative maximum (for $|A_y/A_x|\lesssim 1.3$) or a minimum for higher $|A_y/A_x|$ ratios. For the examples shown in Figs.~\ref{fig1} and~\ref{fig2}, the behavior of the degree of polarization has been varied by changing the $|A_y/A_x|$ ratio, but different $P(\zeta)$ curves can be obtained by changing $q_{xx}$, $q_{yy}$, and $\beta_{xy}$ values.

A straightforward generalization of the previous model can be obtained if one considers a finite sum of the CSDMs of the above type, i.e.,
\begin{equation}
\label{Wn}
\overleftrightarrow{W}^{(0)}(\pmb{\rho}_1,\pmb{\rho}_{2})=\sum\limits_{n=1}^{N}\overleftrightarrow{W}^{(0)}_{n}(\pmb{\rho}_1,\pmb{\rho}_{2}),  
\end{equation}
where the parameters of the individual terms can be chosen for a fine tailoring of $S$ and $P$ of the propagating beam. In this case the elements of $\overleftrightarrow{S}(\zeta)$ are
\begin{equation}
\label{Szn}
\begin{split}
S_{ij}(\zeta)&=\sum\limits_{n=1}^{N}{S_{ij,n}}(\zeta)\\&=\frac{1}{\zeta^2} \sum\limits_{n=1}^{N} \frac{A_{i,n}A_{j,n}\beta_{ij,n}}{q_{ij,n}} \text{Re} \left\{ \widehat{\mu}^{(0)}\left( \frac{\zeta-i}{q_{ij,n} \zeta}  \right)  \right\}, 
\end{split}
\end{equation}
and the on-axis spectral density and degree of polarization become
\begin{equation}\label{SPg}
S(\zeta)= \sum\limits_{n=1}^{N} \text{tr} \overleftrightarrow{S}_n(\zeta), \quad
P(\zeta)=\sqrt{1-\frac{4\text{det} \sum\limits_{n=1}^{N} \overleftrightarrow{S}_n(\zeta)}{\text{tr}^2 \sum\limits_{n=1}^{N} \overleftrightarrow{S}_n(\zeta)}}
\; .  
\end{equation}

\begin{figure}[t]
\centering
\includegraphics[width=0.5\linewidth]{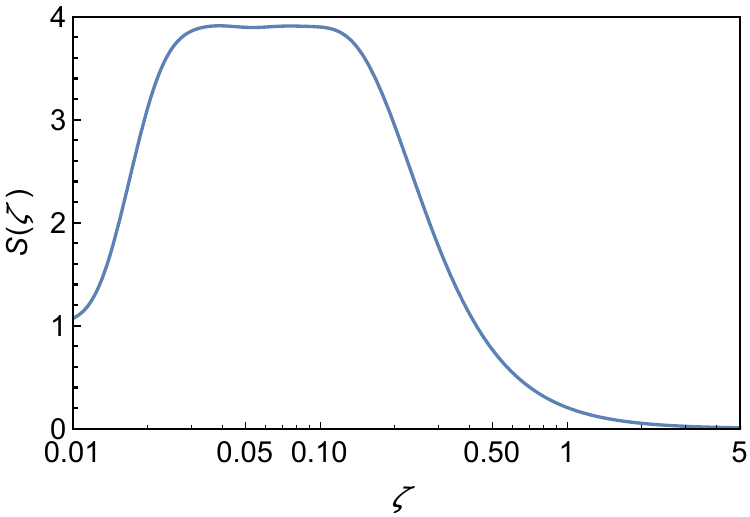}
\caption{$S(\zeta)$ for a beam radiated from a source given by the superposition in \eqref{Wn} with $q_{xx1}$ =2.5, $q_{yy1}$ = 8; $\beta_{xy1}=0$, $A_{x1}=8.2$, $A_{y1}=3.9$; and $q_{xx2}$ =4.5, $q_{yy2}$ = 18; $\beta_{xy2}=0$, $A_{x2}=5.4$, $A_{y2}=3.8$.% and any other permutation of the $x1, y1, x2, y2$ parameters.
}
\label{fig3}
\end{figure}

\begin{figure}[t]
\centering
\includegraphics[width=0.99\linewidth]{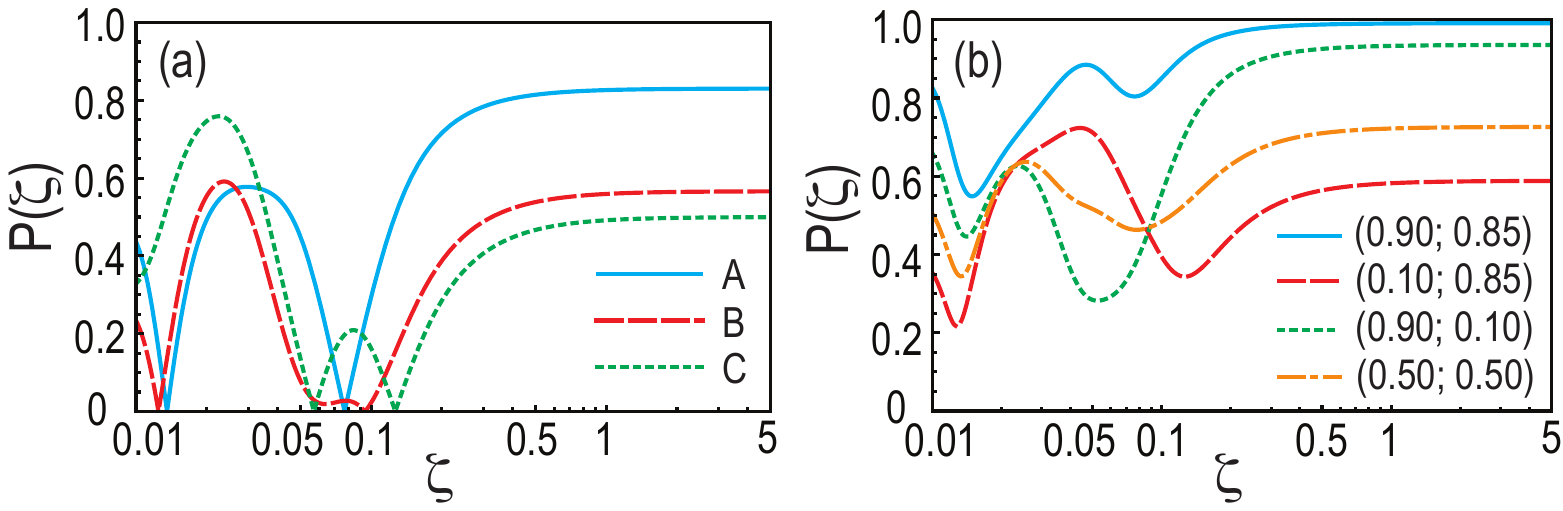}
\caption{$P(\zeta)$ for a beam radiated from a source given by the superposition in \eqref{Wn}  with  $S(\zeta)$ shown in Fig. \ref{fig3}. 
(a): curve A $q_{xx1}$ =2.5, $A_{x1}=8.2$, $q_{yy1}$ = 8, $A_{y1}=3.9$, $\beta_{xy1}=0$; and $q_{xx2}$ =4.5, $A_{x2}=5.4$, $q_{yy2}$ = 18, $A_{y2}=3.8$, $\beta_{xy2}=0$; 
curve B same as curve A, but exchanging $(q_{yy1},A_{y1})$ with $(q_{xx2},A_{x2})$ values;
curve C same as curve A, but exchanging $(q_{xx1},A_{x1})$ with $(q_{yy1},A_{y1})$ values.
(b): $P(\zeta)$ for $q_{xy1}=3.02$, $q_{xy2}=10.2$ and several pairs of $(\beta_{xy1};\beta_{xy2})$ values; the rest of parameters are the same as for curve B in Fig.~\ref{fig4}(a).
}
\label{fig4}
\end{figure}

For instance, Fig.~\ref{fig3} shows that even with $N=2$ it is possible, by suitably choosing the source parameters, to keep $S(\zeta)$ almost constant over a significant propagation range. On the other hand, to one and the same behavior of the on-axis spectral density, $S(\zeta)$, it is possible to associate very different behaviors of the polarization degree $P(\zeta)$, which, in turn, can be finely controlled by tuning the source parameters. This is shown in the curves of Fig. \ref{fig4}. In fact, Fig.~\ref{fig4}(a) shows three curves of $P(\zeta)$ obtained for various combinations of the parameters giving rise to the spectral density of Fig.~\ref{fig3}. In this  case, the curves are obtained for uncorrelated  $x$ and $y$ components ($\beta_{xy1}=\beta_{xy2}=0$), by simply exchanging the values of the parameters for the $x$ and $y$ components of the first and second terms in Eq.~(\ref{Wn}).

The behaviors shown in Fig.~\ref{fig4}(b) correspond to the same parameters values as for the curve B in Fig.~\ref{fig4}(a) but considering several pairs of $\beta_{xy1} \neq 0$ and $\beta_{xy2} \neq 0$ values and choosing $q_{xy1}$ and $q_{xy2}$ within the constrains imposed by Eq.~(\ref{rc_q}). Note that the on-axis degree of polarization behavior is controlled without changing the on-axis spectral density that remains the one shown in Fig.~\ref{fig3}. More complex behavior of the polarization degree can be found by adding more and more terms in the superposition of Eq.~(\ref{Wn}).

\begin{figure}[htbp]
	\centering
	\includegraphics[width=7cm]{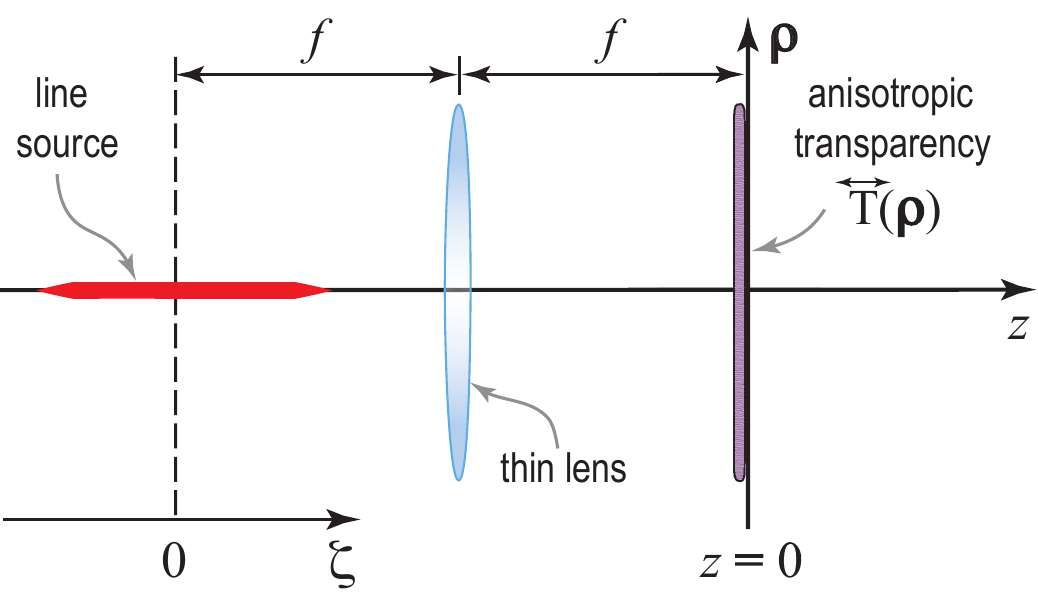}
	\caption{Basic experimental scheme for producing electromagnetic circularly coherent sources.}
	\label{fig5}
\end{figure}

We will now present a possible experimental setup aimed at producing sources of the type introduced in Eq.~(\ref{CSDMcirc}). 
A possible scheme is shown in Fig.~\ref{fig5}. It represents the extension to the vector case of the one that was proposed~\cite{Santarsiero:OL17} and experimentally  tested~\cite{Santarsiero:OL17b} for the scalar case. The source plane is located across the focal plane of a converging lens (focal length $f$). Behind the lens, a line source is placed along the lens axis. A different coordinate ($\xi$, centered at the back focus of the lens) is used to denote a position along the line source. At the plane $z=0$ a planar transparency is placed, able to modify both the amplitude and the polarization of the incident field. It is characterized by the Jones matrix $\overleftrightarrow{T}(\pmb{\rho})$. We assume that paraxial conditions are met for propagation along $z$.

Each element of the source, of length ${\rm d}\xi$, radiates a spherical wave, whose amplitude is proportional to the field $\overrightarrow{E}_s(\xi) {\rm d}\xi$, where each component of $\overrightarrow{E}_s(\xi)$ represents an amplitude density. Such a wave passes through the lens and then through the transparency. 
Starting from the Fresnel propagation formula, it is possible to evaluate the field produced across the plane $z=0$ by a single source element  as 
\begin{equation}
{\rm d}\overrightarrow{E}(\pmb{\rho},\xi) = 
\displaystyle\frac{-{\rm i}}{\lambda f}
\; \overleftrightarrow{T}(\pmb{\rho})
\overrightarrow{E}_s(\xi)
\; e^{{\rm i}k(2 f - \xi)}
\; e^{{\rm i} \; \displaystyle\frac{k}{2f^2} \;  \xi \rho^2}
{\rm d}\xi
\; ,
\label{exp1}
\end{equation}
where $k=2\pi/\lambda$, and the field produced by the whole line source turns out to be
\begin{equation}
\overrightarrow{E}(\pmb{\rho})
= 
\displaystyle\frac{-{\rm i}}{\lambda f}
\; e^{{\rm i}k(2 f - \xi)}
\; \overleftrightarrow{T}(\pmb{\rho})
\displaystyle\int\limits_{-\infty}^{\infty} 
\overrightarrow{E}_s(\xi)
\; e^{{\rm i} \; \displaystyle\frac{k}{2f^2} \;  \xi \rho^2}
{\rm d}\xi 
\; .
\label{exp2}
\end{equation}
Although the source length is generally finite, the integral in Eq.~(\ref{exp2}) was formally extended to ($-\infty,\infty$), the real extension of the segment being considered through function $\overrightarrow{E}_s(\xi)$. 

The corresponding CSDM can be evaluated from Eq.~(\ref{CSDM}) as
\begin{equation}
\begin{array}{l}
\overleftrightarrow{W}^{(0)}(\pmb{\rho}_1,\pmb{\rho}_{2})

\\=
\displaystyle\frac{1}{(\lambda f)^2}
\overleftrightarrow{T}(\pmb{\rho}_1)
\left[
 \displaystyle\int\limits_{-\infty}^{\infty} \int\limits_{-\infty}^{\infty} 
 \overleftrightarrow{W}_s(\xi_1,\xi_2) 
\; e^{{\rm i}k(\xi_2 -\xi_1)}
\right.
\\
\times  
\; e^{\displaystyle - \frac{{\rm i} k}{2f^2} \left( \rho_2^2 \xi_2 - \rho_1^2 \xi_1 \right)}
\; {\rm d}\xi_1 {\rm d}\xi_2
\Bigg]
\; \overleftrightarrow{T}^\dagger(\pmb{\rho}_2)
\; ,
\end{array}
\label{exp4}
\end{equation}
where the CSDM of the field along the source, i.e,
\begin{equation}
\overleftrightarrow{W}_s(\xi_1,\xi_2) 
= \langle 
\overrightarrow{E}_s(\xi_1)\overrightarrow{E}_s^{\dagger}(\xi_2)
\rangle
\; ,
\label{exp5}
\end{equation}
has been introduced.

Equation~(\ref{exp4}) takes a simpler form if the points of the line source are supposed to radiate independently  from one another. In such a case its CSDM takes the form
\begin{equation}
\overleftrightarrow{W}_s(\xi_1,\xi_2) 
= 
\lambda^2 \overleftrightarrow{S}_s(\xi_1)  \; \delta(\xi_1- \xi_2) 
\; ,
\label{exp6}
\end{equation}
where $\overleftrightarrow{S}_s(\xi)$ is the polarization matrix along the line source and $\delta(\cdot)$ is the Dirac's delta function. The expression of the CSDM of the source across $z=0$ then becomes
\begin{equation}
\begin{array}{l}
\overleftrightarrow{W}^{(0)}(\pmb{\rho}_1,\pmb{\rho}_{2})
\\=
\overleftrightarrow{T}(\pmb{\rho}_1)
\left[
\displaystyle\frac{1}{f^2}
\displaystyle\int\limits_{-\infty}^{\infty}
\overleftrightarrow{S}_s(\xi)
\;e^{ - {\rm i} \; {\displaystyle\frac{\pi (\rho_2^2 - \rho_1^2)}{\lambda f^2}} \;  \xi }
\; {\rm d}\xi
\right]
\overleftrightarrow{T}^\dagger(\pmb{\rho}_2)
\; .
\end{array}
\label{exp7}
\end{equation}

The latter CSDM is of the form in Eq.~(\ref{CSDMcirc}), with $\overleftrightarrow{G}^{(0)}$ proportional to the Fourier transform of the polarization matrix $\overleftrightarrow{S}_s(\xi)$ evaluated at the spatial frequency {$\pi (\rho_2^2 - \rho_1^2)/(\lambda f^2)$}. This relation is the same in form as that between $\overleftrightarrow{p}(v)$ and $\overleftrightarrow{G}^{(0)}(\rho_2^2 - \rho_1^2)$ in \eqref{mucirc}. Hence, up to proportionality and scale factors, $\overleftrightarrow{S}_s(\zeta)$ coincides with $\overleftrightarrow{p}(v)$, which, in our case, has Gaussian elements. 
{The $\overleftrightarrow{S}_s(\xi)$ matrix can be generated with the procedure described in reference~\cite{Santarsiero:OL17b}.}
We also remark that the general form of polarization matrix $\overleftrightarrow{T}(\pmb{\rho})$ available in the experimental scheme, is not required for the phenomena we illustrated analytically, and can be used for circularly correlated sources with non-uniform polarization, for example. 

This letter proposes an extension to the vectorial case of the analytic method previously introduced for the scalar case for the study of the axial properties of beams radiated from Gaussian sources with circular coherence. We show that regardless of the source's spatial circular correlations, the on-axis spectral density and the degree of polarization can be found from simple analytic formulas. For illustration of the possibilities in the control of the propagating beam, two examples have been considered. In the first example both the above characteristics are shown to change. In the second example the on-axis spectral density is kept constant within a significant range while the on-axis degree of polarization is shown to be finely controlled by adjusting the source parameters. Thus, the ability to segregate the on-axis dynamics of the spectral density and the polarization properties is demonstrated. % for the first time.
Finally, a simple experimental synthesis scheme is proposed involving a line e.m. source, a converging  lens, and an anisotropic transparency. {The segregation of the on-axis dynamics for the spectral density and the degree of polarization (or, similarly, other polarization properties) can benefit optical systems employing polarization diversity and/or security, such as lidars or free-space communications.}

	\begin{backmatter}
		\bmsection{Funding} Spanish Ministerio de Econom\'ia y Competitividad, project PID2019-104268 GB-C21. OK acknowledges the Copper Fellowship program at the University of Miami.
	
		\bmsection{Disclosures} The authors declare no conflicts of interest.
		
		\bmsection{Data availability} No data were generated or analyzed in the presented research.
		
		%\bmsection{Supplemental document} See Supplement 1 for supporting content. 
		
	\end{backmatter}

	% Bibliography
%	\bibliography{sample}
	
	% Full bibliography added automatically for Optics Letters submissions; the following line will simply be ignored if submitting to other journals.
	% Note that this extra page will not count against page length
%	\bibliographyfullrefs{sample}

\end{document}